\documentclass[%
reprint,
 amsmath,amssymb,
]{revtex4-1}
\usepackage{amsmath}

\usepackage{footnote}

\usepackage{setspace}
\usepackage{lipsum}
\usepackage{graphicx}% Include figure files
\usepackage{dcolumn}% Align table columns on decimal point
\usepackage{bm}% bold math
\usepackage[bottom]{footmisc}
\usepackage{changes}				% Highlight changes
\definechangesauthor[color=red]{MS}
\usepackage[footnote=true]{snotez}	% For comments on changes to show up

\begin{document}

%\preprint{APS/123-QED}

\title{High pressure crystal structure and electronic properties of bismuth silicate Bi\textsubscript{2}SiO\textsubscript{5} from synchrotron experiment and first principle calculations.}% Force line breaks with \\

\author{A. Girard}
\email[Corresponding author: ]{adrien.girard@sorbonne-universite.fr}
\author{M. Stekiel}
\author{W. Morgenroth}
\author{B. Winkler}
\affiliation{Institut f\"ur Geowissenschaften, Goethe Universit\"at Frankfurt, Altenh\"oferallee 1, D-60438 Frankfurt am Main, Germany}%

\author{A. Bosak}
\affiliation{ESRF - The European Synchrotron, 71, Avenue des Martyrs, F-38000 Grenoble, France}%

\author{H. Taniguchi}
\affiliation{Department of Physics, Nagoya University, Nagoya 464-8602, Japan}%

\author{V. Milman}
\affiliation{Dassault Syst\`emes BIOVIA, CB4 0WN Cambridge, United Kingdom}%

\date{\today}% It is always \today, today,

\begin{spacing}{1}

\begin{abstract}
The high pressure structural properties of bismuth oxide
Bi\textsubscript{2}SiO\textsubscript{5} have been investigated up to
28 GPa using \textit{in situ} powder synchrotron X-ray diffraction and
up to 50 GPa with DFT calculations. The monoclinic structure is
found to persist up to $\approx$ 20 GPa, where a notable change in the
compressibility occurs. The DFT data imply that this is due to a
second-order phase transition from the ambient condition monoclinic
structure with space group $Cc$ to an orthorhombic polymorph with
space group $Cmcm$. This transition involves the straightening of the
chains formed by corner-connected SiO$_4$ tetrahedra, that suppresses the ferroelectricity in the high pressure, centrosymmetric phase of Bi$_2$SiO$_5$. The stereo-chemical activity of the Bi$^{3+}$ lone electron pair is found to decrease with increasing pressure, but it can still be identified in the calculated electron density difference maps at 50 GPa.
\end{abstract}

\pacs{Valid PACS appear here}% PACS, the Physics and Astronomy
\maketitle

%\tableofcontents
\section{Introduction}

Currently, there is a considerable interest in ferroelectric lead-free
oxide materials with high transition temperatures. In this context,
Bi\textsubscript{2}SiO\textsubscript{5} (BSO) has emerged as a
promising lead-free oxide, after the finding of high temperature
ferroelectricity with a transition from the $Cc$ to the
$Cmcm$ phase at the Curie temperature $T_C$ = 670 K
\cite{Taniguchi2013,Seol2015}.Apart from ferroelectricity, another interesting electronic feature in BSO is the lone electron pair located at the irregularly coordinated Bi atom. In contrast to other lone-pair-electron-driven displacive-type ferroelectrics \cite{Zhao} or multi- and ferroelectric compounds with perovskite-type structures, where the spontaneous polarization arises from the cation
off-centering (BiFeO$_3$ \cite{Neaton2005},
BiMnO$_3$\cite{Seshadri2001}, PbTiO$_3$), the ferroelectricity in BSO
is primarily driven by the tilt of the SiO$_4$ units \cite{Park2016} within the 1D tetrahedral chains, characteristic for this structure. These findings
have triggered numerous experimental and theoretical studies making
BSO a benchmark system for ferroelectrics based on polymerized
tetrahedra as an alternative to perovskite-type ferroelectrics.

\begin{figure}
	\centering
	\includegraphics[width=0.5\textwidth]{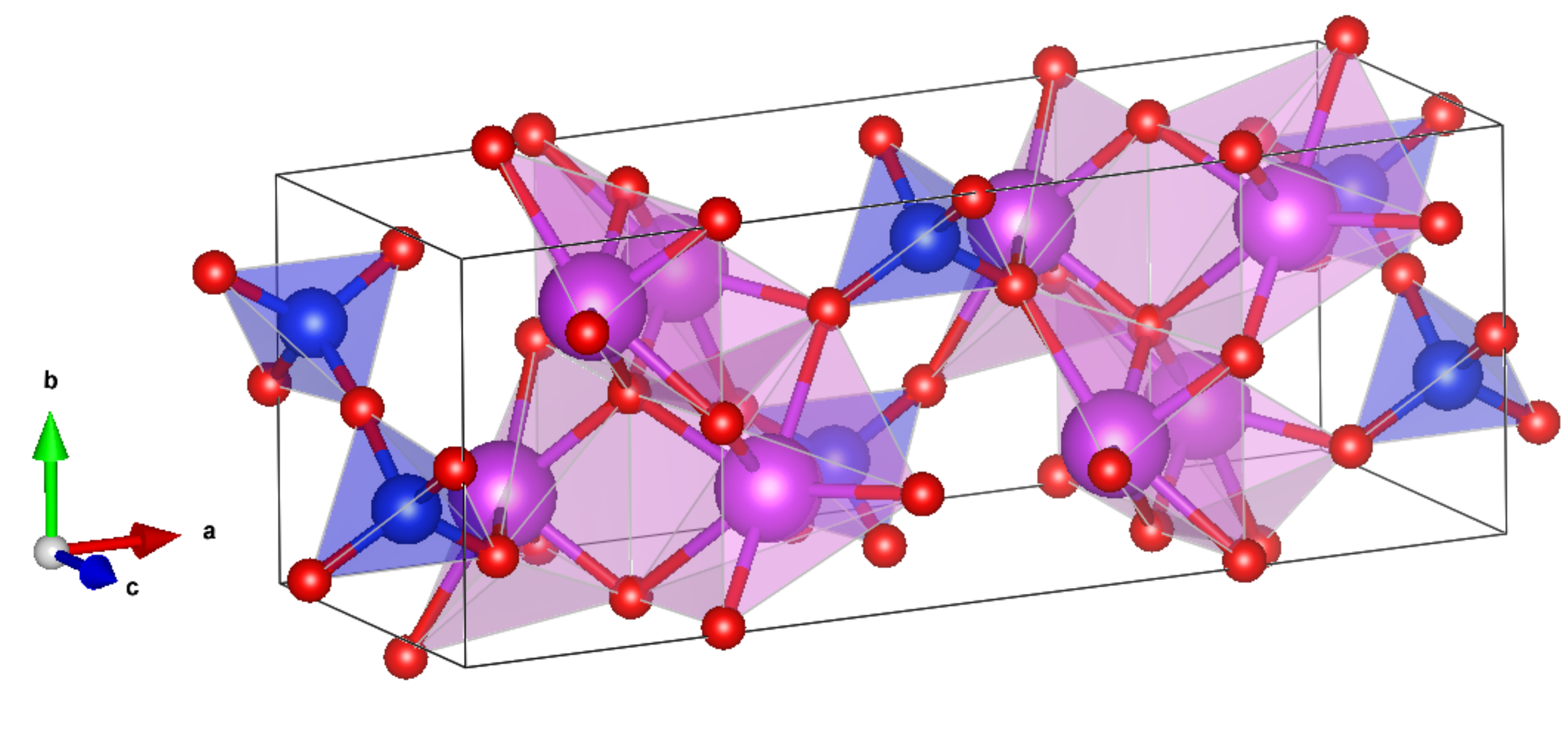}
	\caption{\label{Fig1}Polyhedral representation of the
		unit cell of BSO at ambient pressure conditions. The blue, red and violet
		spheres represent Si, O and Bi atoms, respectively. The blue
		and violet polyhedra indicate the SiO$_4$ and BiO$_6$
		coordination polyhedra. The silicate 1D chains are oriented along the c axis.}
\end{figure}

At ambient conditions BSO is stable in the monoclinic $Cc$ space
group, see structure in Fig.\ref{Fig1}. A very small monoclinic distortion of the crystallographic angle $\beta$ from 90$^{\circ}$ to 90.0695$^{\circ}$ was reported
based on transmission electron microscopy measurements and Rietveld refinements \cite{Taniguchi2013}.  Characteristic for the BSO structure
at ambient conditions are chains of
corner-sharing SiO$_4$-tetrahedra aligned along the
$c$-axis. These chains are located in layers parallel to the
($b,c$)-plane. Bi atoms are also located on planes parallel to the ($b,c$)-plane, and are irregularly coordinated by six oxygens located on one side of Bi atom, while the lone electron pair is located at the other side.  Based on ab initio calculations \citet{Park2016} concluded that the spontaneous polarisation along the $c$-axis was overwhelmingly due to tetrahedral tilting.  A systematic tuning of the ferroelectric phase transition
is possible by element substitution, where Pb was successfully
substituted for Bi up to 20 \% \cite{Taniguchi2016}.

Besides the importance of BSO as a benchmark system to design new
tetrahedra-based ferroelectrics, this compound is also interesting as
a model system to study the interplay between ferroelectric and
antiferroelectric ordering. It was observed by \citet{Taniguchi2013}
that the ferroelectric -- paraelectric phase transition at $T_C$ is
driven by the freezing of an optical phonon at the $\Gamma$ point
\cite{Taniguchi2013}, as is usually observed in ferroelectric
materials \cite{Cochran1960}. However in a recent study an additional
dynamical anomaly at the Brillouin zone edge was reported
\cite{girard2018}, where the Y point optical phonon substantially
softens down to $\sim$50 \% when approaching the $T_C$. The phonon
softening was observed along the whole $\Gamma$ -- Y optical branch,
presumably producing a collapse of the transverse acoustic branch due to TO -- TA mode coupling. Because the Y point phonon is intrinsically associated
with antiferroelectric displacements, the simultaneous observation of
a zone center and a zone boundary dynamical anomaly is evidence for a
competition between ferroelectric and antiferroelectric ordering in
BSO \cite{girard2018}. The outcome of such competition can strongly
depend on an external field, such as pressure. However, the pressure
dependence of BSO has not been reported yet, which was the main
motivation for carrying out the present study.

In this paper we report a high pressure structural study of BSO based
on \textit{in situ} powder diffraction synchrotron experiments and density
functional theory (DFT) calculations. We address the activity of the
stereo-chemically active lone electron pair on the Bi$^{3+}$ ion under
pressure and the compression mechanisms of the ferroelectric $Cc$
structure with an emphasis on the evolution of the 1D tetrahedral
chains responsible for the ferroelectric properties of BSO.

\section{Experimental}
Bi$_2$O$_3$ and SiO$_2$ powders were mixed in an agate mortar in a
stoichiometric ratio and heated in a Pt crucible at 1373 K over 2
h. The resulting polycrystalline melt was subsequently cooled to ambient temperature and crushed to obtain fine powders. 

High pressure diffraction data were recorded at the
Extreme Conditions Beamline P02.2 at PETRA III, Hamburg, Germany.  The
beam was focused to a spot size of 8 $\times$ 2.4 $\mu$m (FWHM, H
$\times$ V) using compound refractive lenses. The incoming photon energy was 42.66 keV (corresponding to 0.29063 \AA) and the detector to sample distance was 400 mm.  

The BSO powder was loaded into the symmetric type diamond-anvil cell (DAC) with 300 $\mu$m culet size diamonds. Rhenium plate pre-indented to 35$\mu$m thickness was used as the gasket, in which a 150 $\mu$m diameter hole was drilled with an excimer laser. Au was used as pressure marker \cite{Jin2011} and neon was employed as a pressure-transmitting medium.

The pressure was increased in steps of 1 to
2 GPa. Before commencing with the data collection we waited for
pressure equilibration.  We collected data at two positions at each
pressure.  The maximum pressure achieved was 29 GPa.  Data collection
was performed in oscillation mode (10$^\circ$ rotation in 10 seconds)
to increase the number of crystallites in diffraction condition. The
two-dimensional XRD images were converted to one-dimensional
diffraction pattern using the DIOPTAS program \cite{Prescher2015}. For
data analysis, Le Bail and Rietveld refinements were carried out with
the Jana2006 software \cite{Petricek2014}.  Additional ambient
pressure data were recorded at the ID28 beamline side station \cite{Sidestation2018} from the
European Synchrotron Radiation Facility (ESRF) with a wavelength
$\lambda$ = 0.5226 \AA.

\section{Computational details}

First-principles calculations were carried out within the framework of
density-functional theory (DFT) \cite{hohenberg64}, the PBE exchange
correlation functional \cite{pbe} and the plane wave/pseudopotential
method using the CASTEP \cite{clark2005} simulation package. "On the
fly" norm-conserving pseudopotentials from the CASTEP data base were
employed in conjunction with plane waves up to a kinetic energy cutoff
of 800 eV. A Monkhorst- Pack [37] grid was used for Brillouin-zone
integrations with a distance of $<$0.023 \AA$^{-1}$ between grid
points. Convergence criteria included an energy change of $<$5
$\times$ 10$^{-6}$ eV/atom for scf-cycles, a maximal force of $<$0.008
eV/\AA , and a maximal component of the stress tensor $<$0.02
GPa. Phonon frequencies were obtained from density functional
perturbation theory-based calculations as implemented in the CASTEP
package.

\section{Results and discussion}

\begin{figure}
	\centering
	\includegraphics[width=0.5\textwidth]{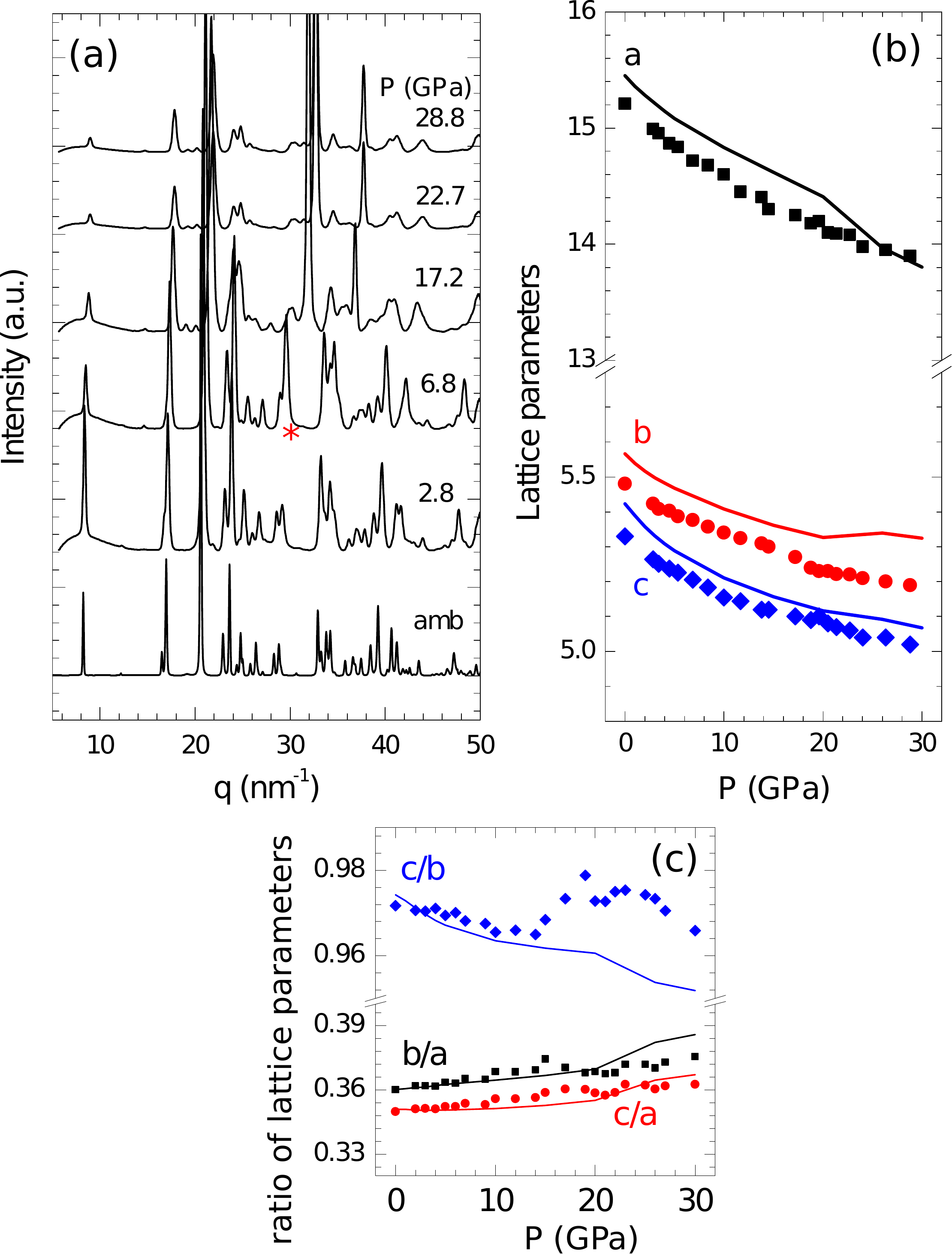}
	\caption{\label{Fig2}
		(a) Selection of XRD patterns of BSO measured at various pressures ($T$ = 300 K, $\lambda$ = 0.2903 \AA). The asterisk
		indicates a diffraction peak due to neon. (b) Pressure dependence of the lattice parameters of BSO. Filled symbols represent experimental data, continuous lines are results from DFT calculations. (c) Ratios of the lattice parameters highlight the change in the compression mechanism
		at  $\approx$ 20 GPa.}
\end{figure} 

\begin{figure}
	\centering
	\includegraphics[width=0.5\textwidth]{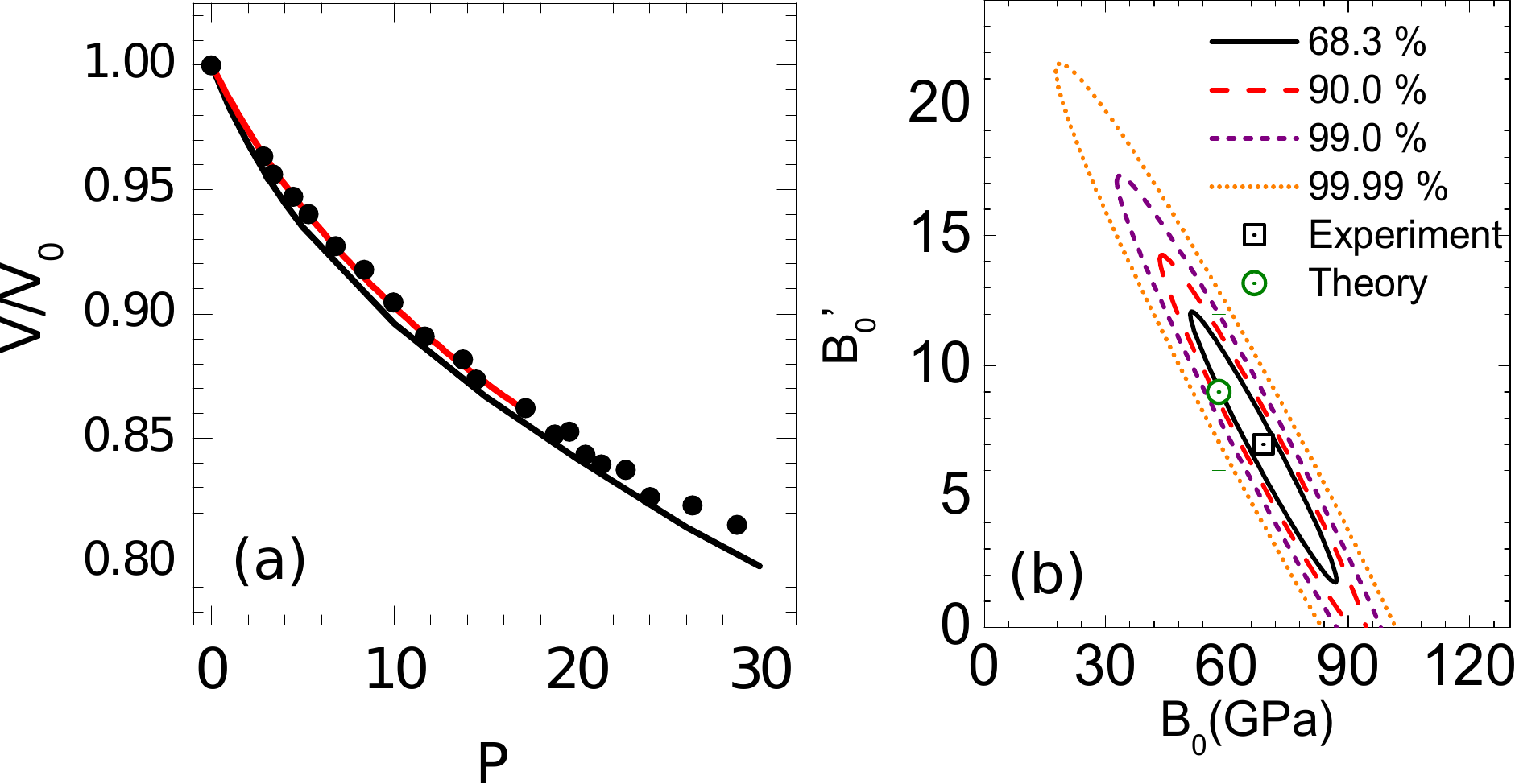}
	\caption{\label{fig:eos} (a) Pressure dependence of the normalized
          unit cell volume of BSO.  Filled symbols represent experimental data, the red line up to 17 GPa is a fit of an equation-of-state to the experimental data and the black line is data from DFT calculations. (b) Bulk modulus ($B_0$) versus bulk
          modulus derivative ($B_0^\prime$) values from the PXRD
          measurement in this study (empty black square) along with
          the DFT value obtained in this study, see
          Table \ref{Bulk_modulus}. Dashed lines present
          confidence ellipses for the fit of $B_0$ and $B_0^\prime$ to
          the measured V(P) relation.}
\end{figure}

A selection of powder diffraction patterns measured at ambient
pressure in a capillary and under pressure up to $P$ = 28.8
GPa in a DAC are shown in Fig.~\ref{Fig2} (a). Due to
the presence of the strongly scattering Bi atoms, the extraction of
reliable positional parameters for the much lighter elements was not
possible for the data collected in a DAC. Hence, the analysis was
restricted to a removal of the background and a subsequent Le Bail
fit, from which the pressure dependence of the lattice parameters were
obtained. The pressure-induced changes are shown in
Fig.~\ref{Fig2} (b). The compression of the $Cc$ structure is
only very slightly anisotropic, with a stronger compressibility of the
long $a$-axis, while the $b$- and $c$-axis show a similar variation
with pressure. The $a$-axis compression is almost linear up to $\sim$
20 GPa while the compressibility decreases above this pressure. Small
changes are also observed at about 19 GPa in the evolution of the $b$-
and $c$-axis, which becomes more obvious by looking the lattice
parameters ratios in Fig.~\ref{Fig2} (c). The ratios $b/a$ and
$c/a$ have a very similar behaviour, slightly increasing up to $\sim$
29 GPa. However, the ratio $c/b$ shows a more interesting pressure
dependence. The $c/b$ ratio continuously decreases up to 17 GPa, where
a kink in the pressure dependence is clearly observed. DFT
calculations show that these compressibility changes are related to
the conformational changes of the chains of Si tetrahedrons in the
($b,c$)-plane and will be discussed in more details below. The P-V data
for the low pressure phase of BSO were fitted with a
third order Birch Murnaghan equation-of-state from ambient pressure up to 17 GPa, see Fig.~\ref{fig:eos} (a). The fit to the experimental P-V data
lead to V$_0= 444.3$ \AA$^3$ (fixed), $B_0=69 (6) $ GPa and $B_0'=7
(1)$, where B$_0$ is the Bulk modulus, B$_0$' its pressure derivative
and V$_0$ the volume of the unit cell at ambient conditions (table
\ref{Bulk_modulus}). The fit parameters of the equation-of-state, namely B$_0$ and
B$_0$', are highly correlated, and to facilitate a comparison with
other data we plotted confidence ellipses in Fig. \ref{fig:eos} (b).

\begin{table}
	\begin{tabular}{c|c|c|c}
                   & exp.        &  DFT           & exp. \\
                   & this study  &  this study    & \cite{Taniguchi2013} \\
  \hline                                         
  a (\AA)          & 15.195(1)      & 15.456         & 15.1193(1)  \\
  b (\AA)          & 5.468(2)       & 5.5632         &  5.4435(1)  \\
  c (\AA)          & 5.315(1)       & 5.4208         &  5.2892(1)  \\
$\beta$ ($^\circ$)  & 90.0        & 89.995         & 90.0695(20) \\
	\end{tabular}
	\caption{\label{latt_param1}Comparison of BSO unit cell parameters
          obtained from DFT calculations, from Lebail refinement
          (this work) and from an earlier study \cite{Taniguchi2013}.}
	
\end{table}

\begin{table}
	\begin{tabular}{c|c|c|c}
		&  B$_0$ (GPa) &  B$_0$' & V$_0$ (\AA$^3$) \\
		\hline
		Experiment & 69 (6) & 7 (1) & 444.3    \\
		DFT from eos & 58 (11) & 9(3) & 466 \\
		DFT from $c_{ij}$ & 55.6(7) &  &  \\
		\hline
	\end{tabular}
	\caption{Bulk moduls B$_0$ (GPa) and its pressure derivative
          B$_0^\prime$ derived from fits of the $P-V$ data for the low
          pressure phase of BSO with a third order Birch-Murnaghan
          equation-of-state from ambient pressure up to 17 GPa, from a
          fit to the theoretical $P-V$ data and from stress-strain
          calculations of the elastic stiffness tensor. }
	\label{Bulk_modulus}
\end{table}

We now turn to our theoretical results. The pressure-dependence of the lattice parameters and their ratios, the equation of state and the resulting bulk modulus obtained from DFT calculations are plotted alongside with the experimental data in Figs.~\ref{Fig2} and \ref{fig:eos}, respectively. Our
experimental and calculated lattice parameters at ambient pressure are
in reasonable agreement ($\sim$ 2\%, see table \ref{latt_param1}) with
each other and earlier data.  The ambient pressure DFT data deviate in
one respect from the structural data published by
\citet{Taniguchi2013}.  In our calculations, the Si-O distances of
oxygens coordinating the Bi atoms are about 1.62 \AA, i.e. a value
typical for Si-O bonds in tetrahedra.  However, the Si-O distances to
those oxygen which are corner shared with the adjacent tetrahedra are
unusually long (1.68 \AA). At ambient pressure the Si-O-Si and O-Si-O angles in the
experimentally determined and computed structure are in very good
agreement. A bond population analysis strongly supports the preference
for a description of the Bi-atom in six-fold coordination, instead of
an alternative description where the Bi-atom is the apex of a pyramid,
as there still is an appreciable bond population of about 0.1
eV/\AA$^3$ even for the longest ($d_{\rm max}$(Bi-O) $\approx$ 2.59
\AA\ Bi-O distances. The irregular coordination polyhedron is,
however, indicative for a stereo-chemical lone electron pair (see
below).

\begin{figure}
	\centering
	\includegraphics[width=0.5\textwidth]{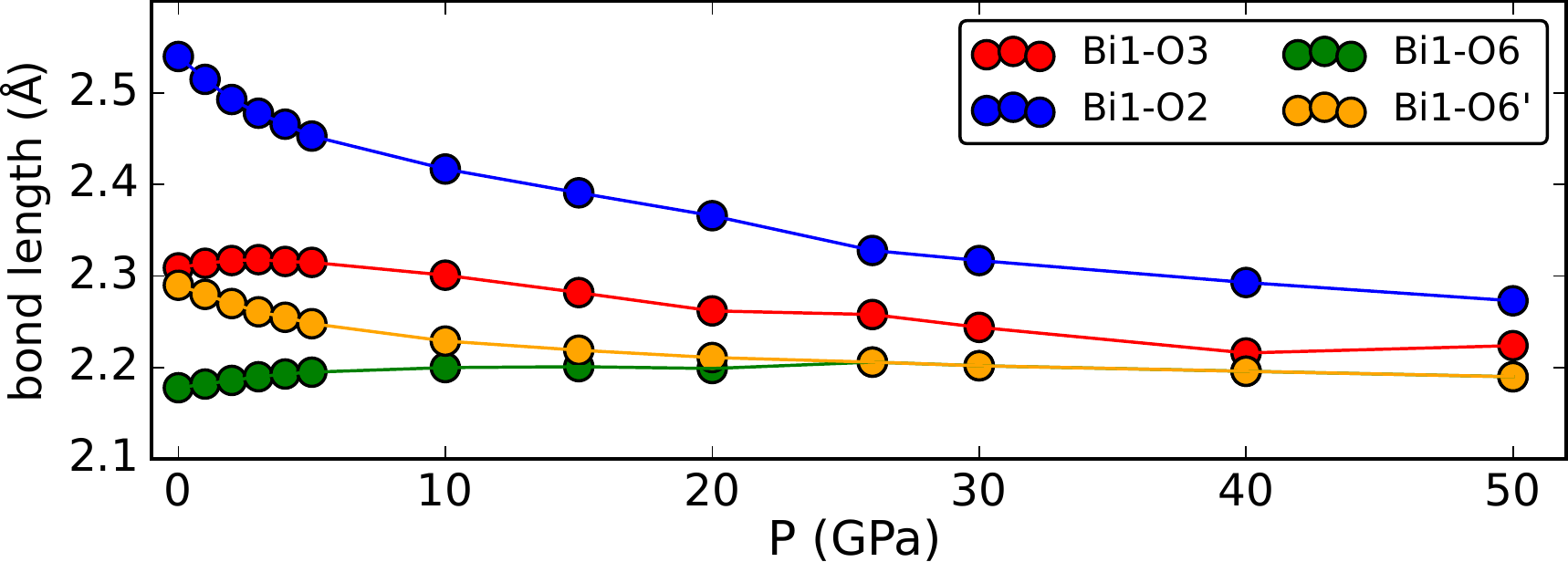}
	\includegraphics[width=0.5\textwidth]{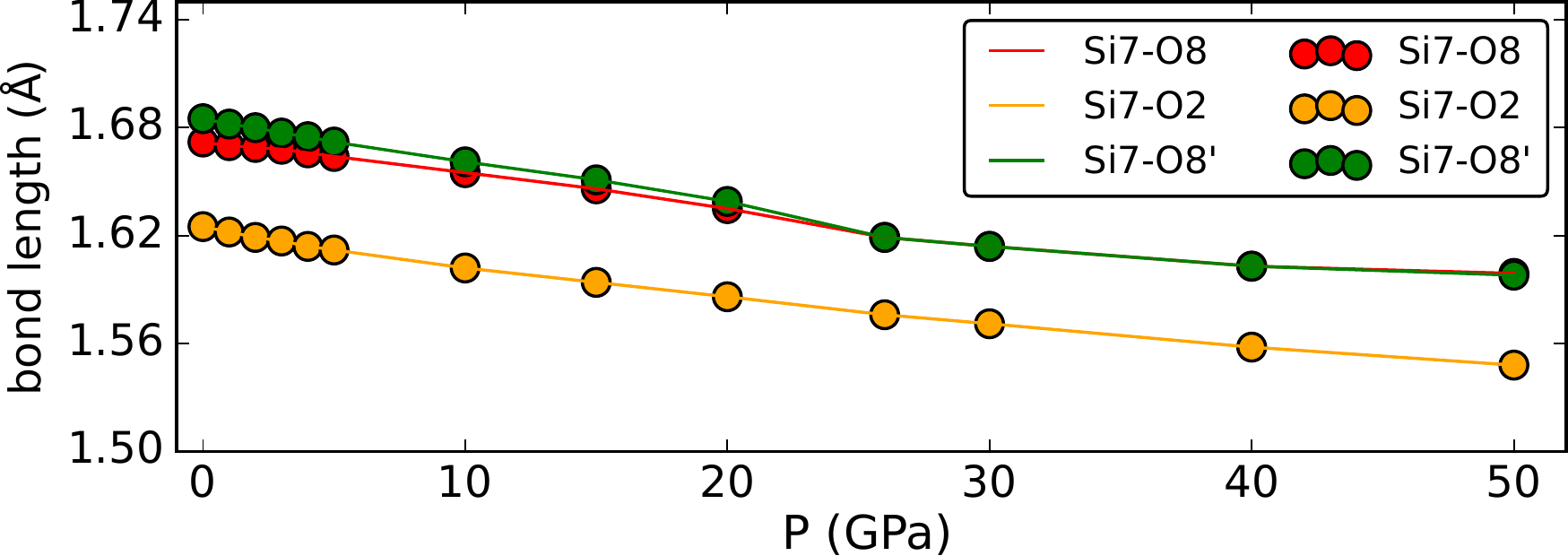}
	\caption{\label{fig:bonds}Pressure dependence of the
		Bi-O and Si-O bond lengths from DFT calculations, which predict
		that on pressure increase the bismuth coordination polyhedra
		becomes less distorted, while in the Si-O tetrahedra two longer and two shorter bonds are preserved up to 50 GPa.}
\end{figure} 

\begin{figure}
	\centering
	\includegraphics[width=0.5\textwidth]{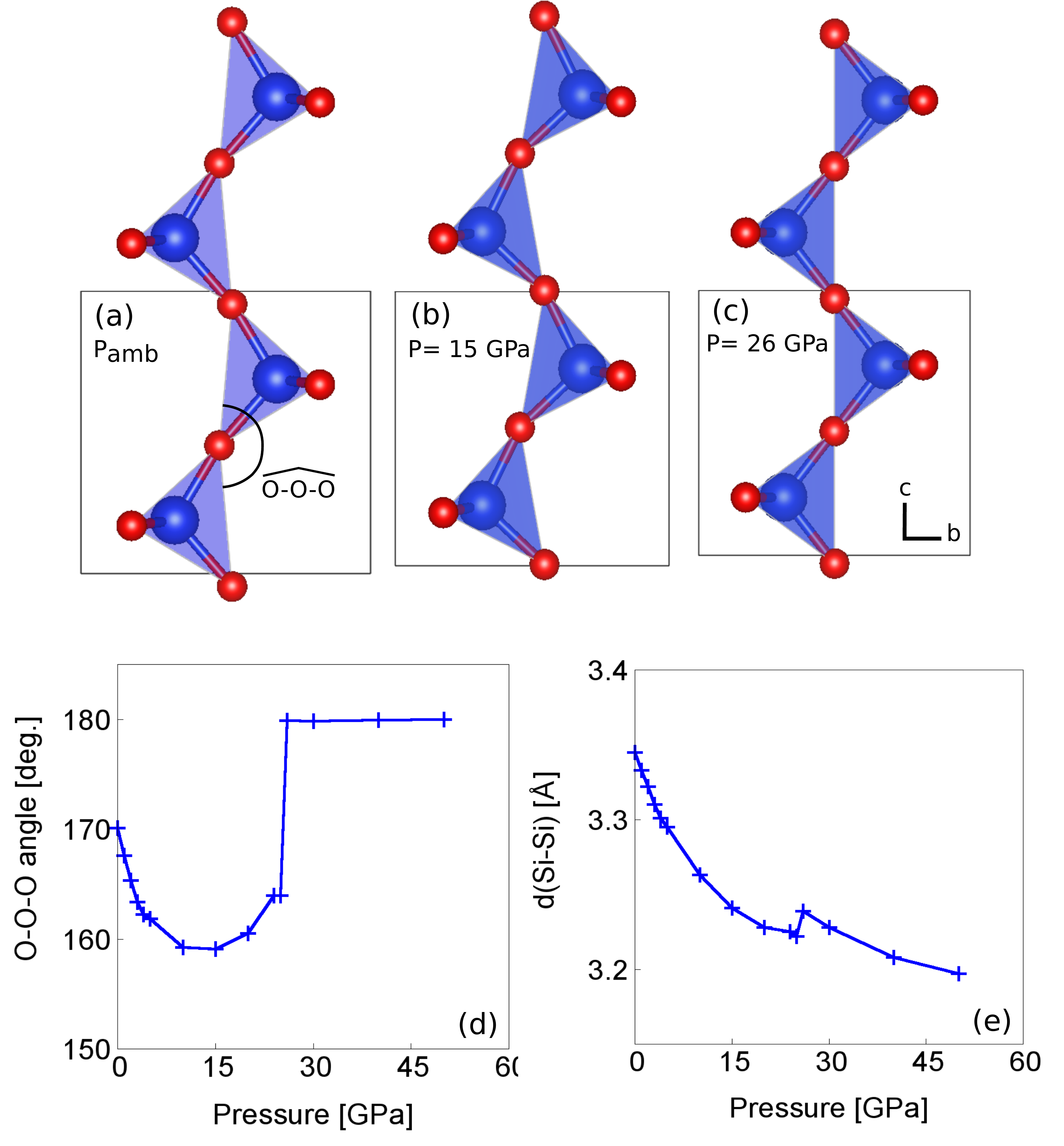}
	\caption{\label{Fig5} (a) - (c) Evolution with pressure of the tilted tetrahedral chains aligned parallel to [001] direction in the ($b,c$)-plane as obtained from DFT. Only the silicate chains are displayed for clarity. (d) and (e) Evolution with pressure of the O-O-O angle and the Si-Si distance, respectively, showing the discontinuity at the phase transition.}
\end{figure} 
The pressure dependence of the lattice parameters obtained from the
DFT calculations shows a similar discontinuous change to that observed
experimentally around 20 GPa, although the effect is slightly more
pronounced in the calculations compared to our observations. 
The DFT model predicted a structural phase transition to an orthorhombic
(space group $Cmcm$) structure, without any discontinuous change in the
unit cell volume.

The fit of an equation of state to the theoreticcal data
(Fig. \ref{fig:eos}) gave $B_0=58 (11) $GPa, $B_0'=9 (3) $ and
V$_0$=466\AA$^3$ (fixed), see table
\ref{Bulk_modulus}. The fit was carried out below the transition pressure. In order to independently assess the bulk
modulus, we carried out stress-strain calculations for the derivation
of the complete elastic stiffness tensor.  We obtained the following
values in GPa: $c_{11}$ = 108(2), $c_{22}$ = 122(2), $c_{33}$ = 131
(1), $c_{44}$ = 42(1), $c_{55}$ = 42.2(5), $c_{66}$ = 48.0 (4),
$c_{12}$ = 45(2), $c_{13}$ = 10(2), $c_{23}$ = 16(2), resulting in a
bulk modulus B$_0$ = 55.6(7) GPa.  All other tensor components which
are allowed to differ from zero in the monoclinic system
($c_{14}$,$c_{15}$,$c_{35}$,$c_{46}$) where zero within the numerical
uncertainty.

The DFT-calculations allow us to follow the pressure-induced structural
changes in detail. The pressure-dependence of the Si-O and Bi-O
bonds are shown in Fig.~\ref{fig:bonds}. As has been noted above, the DFT calculations predict that at ambient conditions there are two bonds with the typical Si-O bond length of 1.62\AA, and two bonds which are slightly longer (1.68 \AA).  At 50
GPa, the bond lenghts have decreased substantially to 1.55 \AA\ and 1.60
\AA, respectively. This prediction is in good agreement with
experimental findings in other silicates.  For example, a single
crystal high pressure diffraction study of a silicate garnet Bi$_2$Ga$_4$O$_9$ gave a Si-O bond length of 1.58 \AA\ at 50 GPa \citep{friedrich2014}. The coordination polyhedron around the Bi becomes less distorted on
increasing pressure. At 50 GPa there are eight oxygen atoms within 2.8
\AA\ of a Bi atom but a population analysis indicates that along all
these distances the population is very low ($<$ 0.08 $e^-$/\AA$^3$
even for the shortest Bi-O distance), and hence there is no
well-defined coordination polyhedron around Bi.

The most striking pressure-induced structural effect in  BSO
is the evolution of the SiO$_4$ tetrahedral chains aligned
along the $c$-axis. The pressure-induced change of the O--O--O angle
formed by the
corner-sharing oxygen atoms along the chain, and the Si-Si distances
are shown in Fig.~\ref{Fig5}. With increasing pressure, the chains shorten, i.e. the Si-Si distances decrease by an increase of the tetrahedral tilt in the chains. Above 20 GPa, a rotation of the SiO$_4$ units leads to straight SiO$_4$
chains. The compressibility of the chain along the chain axis is
significantly smaller for the straight chain than for the
tilted chain. The structural changes are reminiscent of those observed at the ferroelectric -- paraelectric transition at
high temperature \cite{Taniguchi2013}. The Bi -- O and Si -- O bond
lengths are only weakly affected by the rotation. The Si tetrahedral
volume remains constant across the transition. This structural
transition explains the observed anomalies in the compression
data.

\begin{figure}
	\centering
	\includegraphics[width=0.5\textwidth]{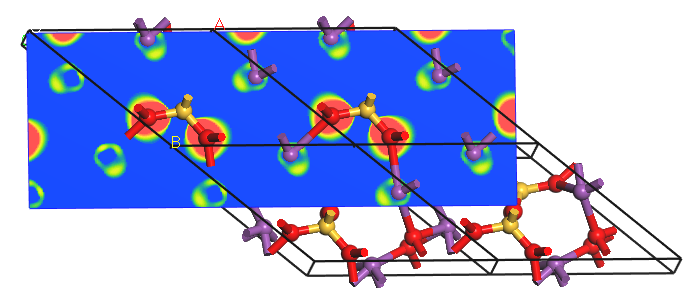}
	\includegraphics[width=0.5\textwidth]{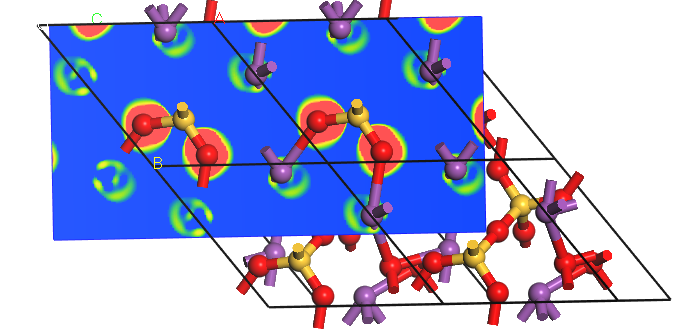}
	\caption{\label{fig:electron} Slice through the calculated
          valence electron density difference function of BSO at ambient
          pressure (top) and 50 GPa (bottom). The maxima of the density difference
          function with a value of $\approx$ 0.04 $e^-$/\AA$^3$ close
          to the Bi-atoms are typical of stereochemically active lone
          electron pairs. While the maxima decrease with
          increasing pressure, they are still discernible at 50 GPa.  }
	
\end{figure} 

We also studied the effect of pressure on the Bi$^{3+}$ 6s$^2$
lone-electron pair. We have evaluated the valence electron density
difference function, i.e. the difference between the overlapping
non-interacting atomic valence electron densities and the
self-consistent electron density up to 50 GPa
(Fig. \ref{fig:electron}). The calculations show clear evidence for a
localised stereochemically active lone electron pair at low pressures,
with a maximum density of $\approx$ 0.04 $e^-$/\AA$^3$ close to the
Bi-atoms. At 50 GPa, this maximum is smeared out, but can still be
detected, and hence the lone electron pair is still stereo-chemically active.
The persistence of stereochemical lone electron pairs at high pressures has been observed in a range of other compounds, including PbO up to 46 GPa \cite{Hausserman2001}, Bi$_2$Ga$_4$O$_9$ \cite{friedrich2010} up to 50 GPa and Bi$_2$S$_3$
\cite{Lundergaard2005,Efthimiopoulos2014}.

As shown by \citet{Park2016} the ferrolectricity of BSO at ambient conditions arises mainly from the tilting of the SiO$_4$ chains forming the SiO$_3$ layer, 24.1 $\mu C/cm^2$, however a small component originates from the Bi$_2$O$_2$ layer, 1.3 $\mu C/cm^2$. They have also shown, that the spontaneous polarization is proportional to the tilting angle, which they defined as an angle complementary to the O--O--O angle defined in this study. This allows us to predict that spontaneous polarization of BSO will increase with pressure, reaching a maximum at $\sim$ 15 GPa with a value $\approx$ 50~$\mu C/cm^2$, two times larger than at ambient pressure. Upon further pressure increase the polarization will decrease, vanishing at $\approx$ 20~GPa due to the transition to the centrosymmetric phase.

\section{Conclusions}
In summary, we have performed a high pressure study of the structure
of BSO by synchrotron X-ray powder diffraction and DFT
calculations. The compression mechanism of the structure below 17 GPa
is characterized by a continuous rotation of the SiO$_4$ tetrahedra,
which leads to a shortening of the Si-Si distances and of the chain
length.  At 17 GPa we observed a structural phase transition during
which the Si tetrahedral chains straighten. The monoclinic angles
differs only very slightly from 90$^\circ$ in the low pressure
monoclinic phase, and only very small changes in the atomic parameters
distinguish the monoclinic phase from the orthorhombic high pressure
phase. This transition is not associated with a change of the
stereochemical activity of the Bi lone electron pair, as this remains
localised up to at least 50 GPa. These results open the route towards dynamical studies of the competition between FE and AFE ordering under high pressure. In particular, the FE soft mode which corresponds to the twisting of the SiO$_4$ tetrahedras, is expected to be strongly affected by the anisotropic compression of the silicate chains.

\section{Acknowledgements}
This study was supported by the BMBF Projects No. 05K16RF1, No. 05K16RF2, No. 05K16RFA, and No. 05K16RFB, and a joint DFG-ANR Project No.WI1232/41-1.
We acknowledge DESY (Hamburg, Germany), a member of the Helmholtz Association HGF, for the provision of experimental facilities. Parts of this research were carried out at PETRA-III and we would like to thank the P02.2 beamline staff for technical support. This work is partially supported by a Grant-in-Aid for Young Scientists (A) (No. 16H06115) and MEXT Element Strategy Initiative Project.

\end{spacing}

%%%REFERENCES%%%
\bibliography{BSO-paper-PRB} 
\bibliographystyle{unsrtnat}

\end{document}